\documentclass[a4paper,11pt]{amsart}
\pdfoutput=1 
\usepackage[normalem]{ulem}

\usepackage{amsmath,amssymb, amsthm, amsfonts, mathrsfs}
\usepackage{xcolor}
\usepackage{graphicx}
\usepackage{marginnote}

\usepackage{hyperref}
\hypersetup{ colorlinks = true,linkcolor = {blue},citecolor = {red} }

\theoremstyle{definition}
\newtheorem{remark}{Remark}
\newtheorem{example}{Example}[section]

\newcommand{\ZZ}{\mathbb{Z}}

\newcommand{\RR}{\mathbb{R}}

\DeclareMathOperator{\Ad}{Ad}
\DeclareMathOperator{\ad}{ad}

\DeclareMathOperator{\Diff}{Diff}
\DeclareMathOperator{\Vir}{Vir}
\DeclareMathOperator{\Vect}{Vect}
\DeclareMathOperator{\Stab}{Stab}

\DeclareMathOperator{\tr}{tr}

\newcommand{\OO}{\mathcal O}

\newcommand{\GG}{\mathcal G}
\newcommand{\g}{\mathfrak g}
\newcommand{\h}{\mathfrak h}

\newcommand{\n}{\mathfrak n}
\newcommand{\Aflat}{\mathcal A_{\rm flat}}
\newcommand{\del}{\partial}

\newcommand{\SL}{{\rm SL}(2,\RR)}
\newcommand{\SLn}{{\rm SL}^{(n)}(2,\RR)}
\newcommand{\PSL}{{\rm PSL}(2,\RR)}

\newcommand{\mat}[4]{ \begin{pmatrix} #1 & #2 \\ #3 & #4 \end{pmatrix} }

\begin{document}

\title[Schwarzian quantum mechanics from $BF$ theory]{Schwarzian quantum mechanics as a Drinfeld-Sokolov reduction of $BF$ theory}

\begin{abstract}
   We give an interpretation of the holographic correspondence between two-dimensional $BF$ theory on the punctured disk with gauge group $\PSL$ and Schwarzian quantum mechanics in terms of a Drinfeld-Sokolov reduction.
   The latter, in turn, is equivalent to the presence of certain edge states imposing a first class constraint on the model.
  The constrained path integral localizes over exceptional Virasoro coadjoint orbits. 
  The reduced theory is governed by the Schwarzian action functional generating a Hamiltonian $S^1$-action on the orbits.
  The partition function is given by a sum over topological sectors (corresponding to the exceptional orbits), each of which is computed by a formal Duistermaat-Heckman integral. 
\end{abstract}

\author{Fridrich Valach}
\address{Department of Physics, Imperial College London, Prince Consort Road, London, SW7 2AZ, United Kingdom \newline \indent Mathematical Institute, Faculty of Mathematics and Physics, Charles University Prague, Prague 186 75, Czech Republic}
\email{f.valach@imperial.ac.uk}

\author{Donald R. Youmans}
\address{Universit\'e de Gen\`eve}
\email{donald.youmans@unige.ch}


\thanks{This work was supported by the NCCR SwissMAP of the Swiss National Science Foundation. F.\ V.\ was further supported by the GA\v{C}R Grant EXPRO 19-28628X and by the Early Postdoc Mobility grant P2GEP2\underline{\phantom{k}}188247 of the Swiss National Science Foundation. Research of D.\ Y.\ was supported by the Grant 178794 of the Swiss National Science Foundation.}

\maketitle

\tableofcontents

\section{Introduction}

Several years ago a correspondence between Jackiw-Teitelboim (JT) gravity on a disk and Schwarzian quantum mechanics on the boundary has been discovered independently in the works of 
 Jensen \cite{Jensen}, Maldacena, Stanford, Yang \cite{Maldacena}, and Engels\"oy, Mertens, Verlinde \cite{Engelsoey-Mertens-Verlinde}.
This is an instance of an ${\rm AdS}_2/{\rm CFT}_1$ correspondence and serves as a toy model for quantum gravity.
Notably, the only dynamical variables of the theory are (orientation preserving) diffeomorphisms of the boundary.
More recently, the partition function of JT gravity on two-dimensional compact surfaces of arbitrary genus and with arbitrary many boundaries has been computed exactly by describing it as a matrix model and employing topological recursion methods \cite{Saad-Shenker-Stanford}. 

JT gravity is a two-dimensional dilatonic gravity theory \cite{Jackiw_1985, Teitelboim}.
It is well-known that the model admits a first order formulation in terms of a two-dimensional $BF$ theory with gauge group $\SL$ \cite{Jackiw_1992}.
In fact, this is a special example of a more general construction of two-dimensional dilatonic gravity theories, which were classified in \cite{Ikeda,Schaller-Strobl} by means of so-called Poisson sigma models.

Schwarzian quantum mechanics is a very rich and interesting theory in itself due to its strong ties with the Sachdev-Ye-Kitaev model \cite{Maldacena-Stanford} (and references therein) where it arises as a low energy limit.
Notably, its partition function is one-loop exact and was calculated in \cite{Stanford-Witten}.
In particular, the path integral is taken over the space of orientation preserving diffeomorphisms $\Diff^{+}(S^1)$ modulo an $\SL$-action.
Now, $\Diff^{+}(S^1)/\SL$ can be identified with an exceptional Virasoro orbit \cite{Alekseev-Shatashvili, Witten} on which the Schwarzian action functional generates a Hamiltonian $S^1$-action by rotating the (source) circle. 
The one-loop exactness of the partition function is therefore a consequence of an analog of the Duistermaat-Heckman integration formula.

Motivated by the above observations, this article is devoted to the study of two-dimensional $BF$ theory on a punctured disk (or cylinder) with gauge group $\PSL$. 
The duality between the Schwarzian theory and $BF$ theory on a disk for gauge group $\SL$ was previously explored in \cite{Astorino-Cacciatori-Klemm-Zanon,Blommaert-Mertens-Verschelde-Wilson_line,Blommaert-Mertens-Verschelde-Fine_structure, Iliesiu,Mertens-Turiaci-Verlinde, Mertens-Turiaci, Saad-Shenker-Stanford}, where it was derived using a combination of holographic methods and Hamiltonian reduction. 

In the present paper we revisit the Hamiltonian reduction. We show that from a mathematical point of view, the constraint can be understood as a Drinfeld-Sokolov reduction for the loop group of $\PSL$. Computations can be explicitly done in the Iwasawa decomposition. This decomposition has the advantage of being global, as opposed to the Gauss decomposition, which is more common in the literature.
The non-trivial topology of the punctured disk allows us to consider loops of arbitrary winding number $n$. The Drinfeld-Sokolov reduction leaves us with loops of positive winding number. In particular, we reobtain the correspondence, discussed in \cite{Mertens-Turiaci}, between winding numbers and exceptional Virasoro coadjoint orbits. Correspondingly, the partition function picks up contributions from all topological sectors. For $n = 1$ we recover results of \cite{Stanford-Witten}.

In more detail: following \cite{Blommaert-Mertens-Verschelde-Wilson_line}, we implement the boundary conditions by adding a Hamiltonian on the boundary.
In addition we assume that half of the fields vanish at the puncture and we restrict the holonomy of the gauge field to be trivial.
After shortly recalling some background material in Section \ref{sec:Kac-Moody-Vir-Orbits}, we recall in Section \ref{sec:BF} how integration over the scalar fields localizes the $BF$ theory over the space of flat connections modulo gauge transformations which are trivial on the boundary.
Restricting the holonomy of the connection around the boundary of the disk to be trivial, this moduli space can be naturally identified with the space of based maps from the boundary to $\PSL$.
Meanwhile, the action functional reduces to quantum mechanics of a free particle moving in the group manifold.
Furthermore, we explain how the presence of edge states, in the sense of \cite{Gegenberg-Kunstatter-Strobl}, constrains the path integral.

A detailed analysis of the constrained model is given in Section \ref{sec:QMonPSL}.
In particular, we explain in detail how the constraint can be understood in terms of a Drinfeld-Sokolov reduction of the space of based loops in $\PSL$.
After the reduction, the path integral localizes to a sum of integrals over exceptional Virasoro coadjoint orbits.
Each orbit is associated to a topologically distinct sector corresponding to different windings of the loop. 
At the same time, the theory reduces to Schwarzian quantum mechanics whose action functional generates an $S^1$-action on the orbits.
To be self-contained, we give an explicit calculation of the partition function in terms of a formal Duistermaat-Heckman integration. 
We find
\[
  Z^{\rm red}(c,\beta) \propto \beta^{-3/2} \sum_{n \geqslant 1}\, n\, \exp\left( -\frac{\pi c n^2}{12 \beta \hbar} \right),
\] 
which recovers the results of \cite{Stanford-Witten} for $n=1$.

Furthermore, we show that in the presence of a fixed edge state the constrained partition function reduces to a sum of integrals over non-exceptional Virasoro coadjoint orbits.
Again, the reduced action generates a Hamiltonian $S^1$-action.
Once more, we compute the partition function by means of a formal Duistermaat-Heckman integration, recovering again the results of \cite{Stanford-Witten}. 


Let us stress that although the present work is motivated by 2D quantum gravity (and the corresponding holography), our main interest is not in gravity but in the geometric understanding of the emergence of the Schwarzian theory at the boundary via a Drinfeld-Sokolov reduction and the appearance of higher modes.\bigskip

\noindent\textbf{Acknowledgements.}\! The authors are grateful to Anton Alekseev and Samson Shatashvili for many insightful discussions and useful comments. They also thank the anonymous referee for valuable remarks and suggestions which helped to improve the article.

\section{Preliminaries}\label{sec:Kac-Moody-Vir-Orbits}
\subsection{Finite-dimensional Duistermaat-Heckman integration}\label{sec:DH}
Let us start with a non-exhaustive recollection of finite-dimensional Duistermaat-Heckman integration.
For more details we refer the interested reader to \cite{Alekseev,Atiyah-Bott,Picken}. 

Let $(M,\omega)$ be a compact symplectic $2n$-dimensional manifold endowed with an action of $S^1$.
Suppose that this circle action is Hamiltonian, that is we assume that the action is generated by a vector field $\xi$ and the existence of a smooth function $H$ on $M$ which satisfy the relation
\[
  \iota_{\xi} \omega + dH = 0. 
\]
Moreover, suppose that $H$ has only isolated critical points.
Then, Duistermaat and Heckman showed in \cite{Duistermaat-Heckman} that the integral
\[
  I(\varepsilon) = \int_M e^{-\varepsilon H} \frac{\omega^n}{n!}
\]
localizes over the fixed points of $H$: 
\begin{equation}
  I(\varepsilon) = \sum_{m} \frac{e^{-\varepsilon H(m)}}{\prod_{j=1}^n \tfrac{\varepsilon}{2\pi} w_j(m)},
  \label{eq:DH_integration_formula}
\end{equation}
where the sum runs over all (isolated) critical points $m$ of $H$ and the $w_j(m)$ are the weights of the $S^1$-action on the tangent space $T_mM$ of $M$ at $m$.
Finally, the integration measure is taken to be the Liouville measure $\omega^n/n!$ defined by the symplectic form $\omega$.
\subsection{Kac-Moody orbits}
Let $G$ be a semisimple Lie group with Lie algebra $\g$.
Denote by $LG = C^{\infty}(S^1,G)$ the loop space of $G$. 
The space $LG$ is itself a Lie group, whose Lie algebra $L\g$ coincides with the algebra of smooth $\g$-valued functions on the circle.

In the following, we will be interested in the central extension $\widehat{L\g}$.
Elements of $\widehat{L\g}$ are of the form $(u(x),k)$, where $u(x)$ is a $\g$-valued function of $S^1$ and $k \in \RR$ is central.
The Lie bracket on $\widehat{L\g}$ is defined by 
\begin{equation}
  [(u,k),(v,\ell)] = \left( [u,v]_{\g}, \frac{1}{2\pi} \tr \oint v(x) u'(x) dx \right)
\end{equation}
where $[u,v]_{\g}$ denotes the bracket in the Lie algebra $\g$ and $\tr$ denotes the normalized Killing form on $\g$.
Elements of the dual space $\widehat{L\g^*}$ can be described equally by pairs $(v(x),-k)$, where $v(x)$ is again $\g$-valued function on $S^1$ and $k$ a real number. 
The number $k$ is called the level. 

The pairing between $\widehat{L\g^*}$ and $\widehat{L\g}$ is 
\begin{equation}
  \left\langle (v,-k), (u,\ell) \right\rangle = \tr \oint v(x)u(x) dx - k\ell.
\end{equation}
The coadjoint action of $LG$ on $\widehat{L\g^*}$ is then
\begin{equation}
  (v,-k)_g \equiv \Ad^*_g (v,-k) = \left( g v g^{-1} + \frac{k}{2\pi} g'g^{-1}, -k  \right).
  \label{eq:Kac-Moody_coadj}
\end{equation}
The coadjoint orbit\footnote{For the sake of readability, we will often refrain from writing the explicit dependence on the level, e.g.\ we write $\OO_{v}$ instead of $\OO_{(v,-k)}$ and $\Stab(v)$ instead of $\Stab( (v,-k) )$.} $\OO_{v} \subset \widehat{L\g^*}$ passing through $(v,-k)$ is the set of all $(v,-k)_g$ for $g \in G$, and is isomorphic to the homogeneous space $LG / \Stab(v)$.

Since any coadjoint orbit is naturally symplectic, so is $\OO_v$.
The symplectic form on $\OO_v$ descends from the left-invariant pre-symplectic form on $LG$, which can be described as follows: 
For $X,Y \in T_e(LG) = L\g$, one defines 
\begin{equation}
  \omega_e(X,Y) = \left\langle (v,-k), [(X,0),(Y,0)] \right\rangle = \tr \oint \left( v[X,Y]_{\g} + \frac{k}{2\pi} XY'  \right)dx.
\end{equation}
Using left translation in $LG$, the pre-symplectic form can then be defined at any $g \in LG$:
\begin{equation}
  \omega_g = \tr \oint \left( v (g^{-1}\delta g)^2 + \frac{k}{4\pi} g^{-1}\delta g \wedge (g^{-1}\delta g)' \right)dx,
\end{equation}
where $g^{-1} \delta g$ denotes the left-invariant Maurer-Cartan element of $LG$.

It is instructive to look at two examples in detail.
\begin{example}\label{ex:OmegaG}
Suppose that $v = 0 \in L\g^*$.
Then the stabilizer $\Stab(v)$ is the group $G$ itself and the coadjoint orbit passing through $0$ is isomorphic to the space of based loops 
\[
  \OO_0 \cong LG/G \cong \Omega G = \{ g \in LG \mid g(0) = e \in G \}.
\]
The symplectic form, at level $k$, reduces to
\begin{equation}
  \omega_g = \frac{k}{4\pi} \tr \oint \left( g^{-1}\delta g \wedge (g^{-1}\delta g)'  \right)dx.
  \label{eq:OmegaG_omega}
\end{equation}
Rotating the loop defines an $S^1$-action on $\Omega G$ by:\footnote{Here, and henceforth, we take $S^1 \cong \RR/2\pi\ZZ$. In particular, we will use the additive notation for the group structure on $S^1$.} $g(x) \mapsto g(x + t)$, $t \in S^1$. 
This action is Hamiltonian, see e.g.\ \cite{Atiyah-Bott,Pressley-Segal}.
Its generating Hamiltonian, with respect to the symplectic form~\eqref{eq:OmegaG_omega}, is the energy function $H \colon \Omega G \to \RR$ of the loop defined by 
\begin{equation}
  H(g) = -\frac{k}{4\pi} \tr \oint (g'g^{-1})^2dx. 
  \label{eq:H_Omega_G}
\end{equation}

\end{example}

\begin{example}\label{ex:LG/K}
Let $T \subset G$ be a maximal torus with Lie algebra $\mathfrak t$, and consider a constant regular element $v_0 \in \mathfrak{t}^* \subset L\g^*$.
Here, \emph{regular} means that the stabilizer subgroup of $v_0$ is a maximal torus. 
Then the orbit through $v_0$ is isomorphic to the homogeneous space $LG / T$. 
In this case, the symplectic form can be conveniently written in terms of the quasi-periodic element $h(x) = g(x) \exp(\tfrac{2\pi}{k} v_0 x)$:
\begin{equation}
  \omega_g = \frac{k}{4\pi} \tr \oint h^{-1}\delta h \wedge (h^{-1}\delta h)' dx.
\end{equation}
The $S^1$-action which rotates the loop is again Hamiltonian, with
\begin{equation}
  H(g) = -\frac{k}{4\pi} \tr \oint (h'h^{-1})^2 dx = -\frac{k}{4\pi} \tr \oint \left(\frac{2\pi}{k}v_0 + g^{-1}g'\right)^2 dx.
  \label{eq:H_Kac-Moody}
\end{equation}
\end{example}

\subsection{Virasoro orbits}\label{sec:Virasoro_orbits}
This section follows closely \cite{Alekseev-Shatashvili, Witten}.
Let $\Vect(S^1)$ be the Lie algebra of the orientation preserving diffeomorphism group of the circle, denoted by $\Diff^{+}(S^1)$.
Elements $\phi\in\Diff^{+}(S^1)$ can be seen as increasing quasi-periodic maps, i.e.\ $\phi \colon \RR \to \RR$ satisfying 
\[
  \phi(x + 2\pi) = \phi(x) + 2\pi.
\]

Let $\Vir$ be the central extension of $\Vect(S^1)$ by $\RR$.
Elements of $\Vir$ are pairs $(v,r)$ where $v = v(x)\del_x$ is a vector field on $S^1$ and $r \in \RR$ is central.
The Lie bracket on $\Vir$ is defined by
\begin{equation}
  [(v_1,r_1), (v_2,r_2)] = \left( [v_1,v_2]_{\Vect(S^1)}, \frac{1}{48\pi} \oint v'''_1 v_2 - v_1 v'''_2  \right),
\end{equation}
where $[v_1,v_2]_{\Vect(S^1)} = (v_1v'_2 - v'_1v_2)\del_x$ stands for the Lie bracket in $\Vect(S^1)$.
Usually one defines elements $L_n$ of the complexification of $\Vir$ by
\begin{equation}
  L_m = i e^{imx}\del_x.
  \label{eq:Vir_generators}
\end{equation}
Then the corresponding commutators
\begin{equation}
  [(L_m,a),(L_n,b)] = \left( (m-n)L_{m+n}, \frac{m^3}{12} \delta_{m+n,0} \right) 
\end{equation}
define the Virasoro algebra relation, which, in the literature, is more commonly written as
\begin{equation}
  [L_m,L_n] = (m-n)L_{m+n} + \frac{c}{12}m^3 \delta_{m+n,0}.
\end{equation}
\begin{remark}
  Shifting $L_0$ by $\frac{c}{24}$ 
  amounts to replacing $m^3$ by $m^3 - m$, which is found more often in the literature.
\end{remark}

However, we will be interested in the real space $\Vir$, which is spanned by the linear combinations
\[L^+_n=-\tfrac{i}{2}(L_n+L_{-n})=\cos(nx)\partial_x,\qquad L^-_n=\tfrac12(L_n-L_{-n})=\sin(nx)\partial_x,\]
for $n>0$, together with $\hat L_0=-iL_0=\partial_x$ and the central element. In particular, since
\begin{equation}
\label{eq:action_L0_on_Lpm}
[\hat L_0,L^\pm_n]=\pm n L^\mp_n,
\end{equation}
the group $S^1$ generated by $\hat L_0$ acts with weight $n$ on the two-dimensional subspace spanned by $L^\pm_n$.

Elements of the dual space $\Vir^*$ are pairs $(b,c)$ where $b = b(x)dx^2$ is a quadratic differential on the circle and $c \in \RR$. 
The pairing between $\Vir$ and $\Vir^*$ is given by
\begin{equation}
  \left\langle (b,c), (v,d) \right\rangle = \oint b(x)v(x)dx + cd.
\end{equation}
The coadjoint action of $\Diff^{+}(S^1)$ on $\Vir^*$ is described as follows: infinitesimally, i.e.\ for $v \in \Vect(S^1)$, one has
\begin{equation}
  \ad^*_{v} (b,c) = \left( (2 v'(x) b(x) + v(x) b'(x)  - \frac{c}{24\pi} v'''(x))dx^2 , 0 \right).
  \label{eq:inf_coad}
\end{equation}
This integrates to
\begin{equation}
  (b,c)_{\phi} \equiv \Ad^*_{\phi^{-1}} (b,c) = \left( \phi^* b - \frac{c}{24 \pi} \{ \phi, x \} dx^2, c \right)  
  \label{eq:Vir_coad}
\end{equation}
where $\phi \in \Diff^{+}(S^1)$ and
\begin{equation}
   \{ \phi,x \}  = \frac{\phi'''}{\phi'} - \frac{3}{2} \left( \frac{\phi''}{\phi'} \right)^2 = \left( \frac{\phi''}{\phi'}  \right)' - \frac{1}{2} \left( \frac{\phi''}{\phi'} \right)^2
  \label{eq:Schwarzian}
\end{equation} 
denotes the Schwarzian derivative of $\phi$.

We can identify the coadjoint orbit $\OO_{b}$ through the point $(b,c) \in \Vir^*$ with
\begin{equation}
  \OO_b \cong \Diff^{+}(S^1) / \Stab(b)
\end{equation}
where $\Stab(b) \subset \Diff^{+}(S^1)$ is the stabilizer subgroup of $b$ under the coadjoint action~\eqref{eq:Vir_coad}.

The stabilizers of a general quadratic differential $b \in \Vir^*$ are hard to calculate.
However, if $b = b_0 dx^2$, $b_0 \in \RR$ is constant, the calculation becomes feasible: A point in the orbit\footnote{By abuse of notation, we will often write $b_0$ instead of $b_0dx^2$.} 
$\OO_{b_0}$ can be written as
\begin{equation}
  (b_0,c)_{\phi} =  \left( \Big( \phi'^2 b_0  - \frac{c}{24 \pi} \{ \phi,x \}  \Big) dx^2, c \right).
\end{equation}
For generic $b_0 \in \RR$, $(b_0,c)$ is only invariant under a (constant) translation, namely $\phi(x) = x + a$. 
This shows that $\Stab(b_0) = S^1$ such that the orbit is isomorphic to the homogeneous space $\Diff(S^1) / S^1$.

However, for exceptional values of $b_0$ one finds that the stabilizer subgroup of $b_0$ is larger than $S^1$: choosing the vector field $v$ to be $L^\pm_n$ in Equation~\eqref{eq:inf_coad}, one has for constant $b_0$ 
\begin{equation}
  \begin{split}
    ad^*_{L^+_n} (b_0,c) &= \left( -2 n \sin(nx) \left( b_0 + \tfrac{cn^2}{48\pi} \right)dx^2, 0 \right), \\
    ad^*_{L^-_n} (b_0,c) &= \left(\phantom{-}2 n \cos(nx) \left( b_0 + \tfrac{cn^2}{48\pi} \right)dx^2, 0 \right).
\end{split}
\end{equation}
For the special values
\begin{equation}
  b_0 = -\frac{cn^2}{48\pi},
  \label{eq:exceptional_b}
\end{equation}
the stabilizer is generated by $\{ \hat L_0, L^\pm_{n} \}$, which integrates to the subgroup $\SLn \subset \Diff^+(S^1)$, the $n$-fold cover of $\SL$ \cite{Witten}.
Thus in this case the orbits are identified with the homogeneous space $\Diff^{+}(S^1) / \SLn$:
\begin{equation}
  \OO_n \cong \Diff^{+}(S^1) / \SLn.
  \label{eq:orbit_n}
\end{equation}
Being a coadjoint orbit, $\OO_{b_0}$ is naturally symplectic. 
In terms of the left-invariant Maurer-Cartan element $Y$ of $\Diff^{+}(S^1)$ \cite{Alekseev-Shatashvili}, the symplectic form at any point $\phi$ is given by 
\begin{equation}
  \omega_{\phi} = \left\langle (b_0,c)_{\phi}, [Y(\phi),Y(\phi)] \right\rangle.
  \label{eq:Omega}
\end{equation}
Explicitly, we have $Y(\phi) = \tfrac{\delta \phi}{\phi'}$, where $\delta$ denotes the de Rham differential on $\Diff^{+}(S^1)$. 
Then, for $b_0 = -\tfrac{c b^2}{48 \pi}$ with $b \in \RR$, one finds (c.f. \cite{Alekseev-Shatashvili, Stanford-Witten}) 
\begin{equation}
    \omega_{\phi} = - \frac{c}{48\pi} \oint \left( b^2 \delta \phi \wedge \delta \phi' - \frac{\delta \phi' \wedge \delta \phi''}{\phi'^2} \right) dx.
  \label{eq:Vir_omega}
\end{equation}

In the following, we will be interested in an infinite-dimensional version of Duistermaat-Heckman integration over the orbits $\OO_{b_0}$.
To this end, let us remark that there exists again an $S^1$-action on $\OO_{b_0}$ which rotates the source circle: $\phi(x) \mapsto \phi(x + a)$.
This $S^1$-action is Hamiltonian with respect to the symplectic form~\eqref{eq:Vir_omega}, with Hamiltonian 
\begin{equation}
  H(\phi) = \frac{c}{24\pi} \oint \left( \frac{b^2}{2} \phi'^2 + \{ \phi, x \}  \right)dx.
  \label{eq:Vir_Hamiltonian}
\end{equation}

\subsection{Drinfeld-Sokolov reduction}\label{subsec:DS}
Let $(M, \omega)$ be a symplectic manifold endowed with a Hamiltonian action of a Lie group $G$.
Let furthermore $\xi\in \g^*$ be a regular value of the moment map $\mu\colon M\to \g^*$. We can then consider the \emph{symplectic reduction}
\[M_{\rm red} = M /\!\!/ G := \mu^{-1}(\xi) / \Stab(\xi).\] 
This is naturally a symplectic manifold, with $\omega_{\rm red}$ given by $\iota^* \omega  = \pi^* \omega_{\rm red}$, where $\iota \colon \mu^{-1}(\xi) \hookrightarrow M$ and $\pi \colon \mu^{-1}(\xi) \twoheadrightarrow M_{\rm red}$ \cite{Marsden-Weinstein}.

Now, the orbits of the coadjoint action of $G$ on $\g^*$ are canonically symplectic. 
Let $\OO\subset \g^*$ be such an orbit and $\mathfrak{h}\subset \g$ be a Lie subalgebra, corresponding to a closed Lie subgroup $H \subset G$. Then the coadjoint action of $H$ on $\OO$ is Hamiltonian, with moment map given by the projection from $\OO\subset \g^*$ to $\h^*$. Thus, given an orbit $\OO \subset \g^*$, $\h\subset \g$ and a suitable $\xi\in \h^*$, we can consider the symplectic reduction $\OO/\!\!/H = \mu^{-1}(\xi)/\Stab(\xi)$.

In the special case when $\g$ is semisimple and $\h$ is nilpotent, the symplectic reduction is known as \emph{Drinfeld-Sokolov reduction}.
Originally, this construction was introduced and studied by Drinfeld and Sokolov in \cite{Drinfeld-Sokolov}, where they used it to derive the bi-Hamiltonian structure underlying the KdV equation by studying the symplectic reduction of Kac-Moody coadjoint orbits.

\section{\texorpdfstring{$BF$}{} theory on a punctured disk}\label{sec:BF}

\subsection{Quantum mechanics in \texorpdfstring{$\PSL$}{} as a holographic dual}
Let $D^* = D - \{ 0 \}$ be the punctured unit disk and let $G = \PSL$.
We denote by $\g$ its Lie algebra and by $\tr$ the non-degenerate Killing form.
The $BF$ theory we are interested in is defined by a $\g$-valued scalar field $X$ and a $\g$-valued one form\footnote{It is customary to think of $A$ as a connection of a trivial principle $G$-bundle over $D^*$.} $A$.
The action functional of the model is
\[
  S(X,A) = \int_{D^*}  \tr X F_A + \oint_{\del D^*} \tr XA, 
  \label{eq:BF}
\] 
where $F_A = dA + \tfrac{1}{2}[A,A]$ denotes the curvature of $A$.
In order to have a well-defined Dirichlet problem for the variational principle, we need to specify boundary conditions. 
Following \cite{Blommaert-Mertens-Verschelde-Wilson_line}, we choose to implement these boundary conditions by adding a Hamiltonian on the boundary:
\begin{equation}
  S(X,A) = \int_{D^*}  \tr X F_A + \oint_{\del D^*} \tr XA - \frac{1}{2}\tr X^2 dx,
  \label{eq:first_order}
\end{equation}
where $dx$ denotes a volume form on $S^1 = \del D^*$ and the Hamiltonian $\tr X^2$ is, up to a constant, the quadratic Casimir of $\PSL$.
Moreover, we assume that the scalar field $X$ vanishes at the puncture.
\begin{remark}
  We can think of the punctured disk as a semi-infinite cylinder $(-\infty,t_0) \times S^1$. 
  Let $t \in (-\infty,t_0)$ be the coordinate along the cylinder and $x$ the angle coordinate.
  Then the origin, i.e.\ the puncture of the disk corresponds to $t \to -\infty$.
  The assumption that $X$ vanishes at the puncture of the disk is therefore equivalent to the assumption that $X$ vanishes at infinity.
\end{remark}

\begin{figure}[htb]
  \centering
  \includegraphics[scale=.6]{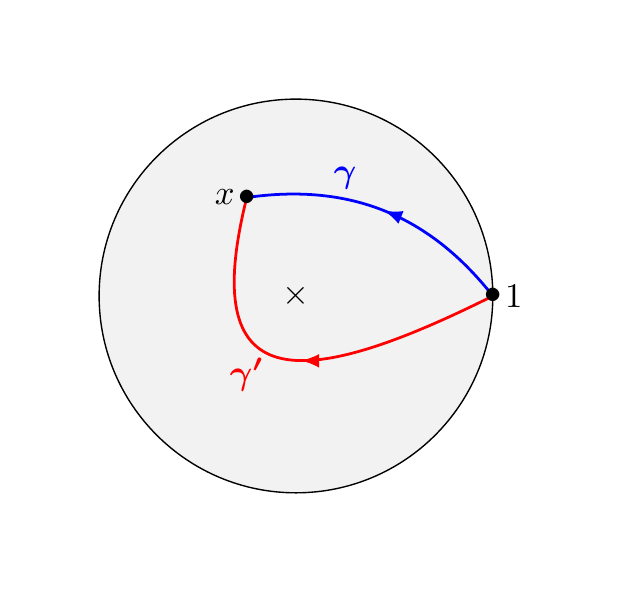}
  \includegraphics[scale=.6]{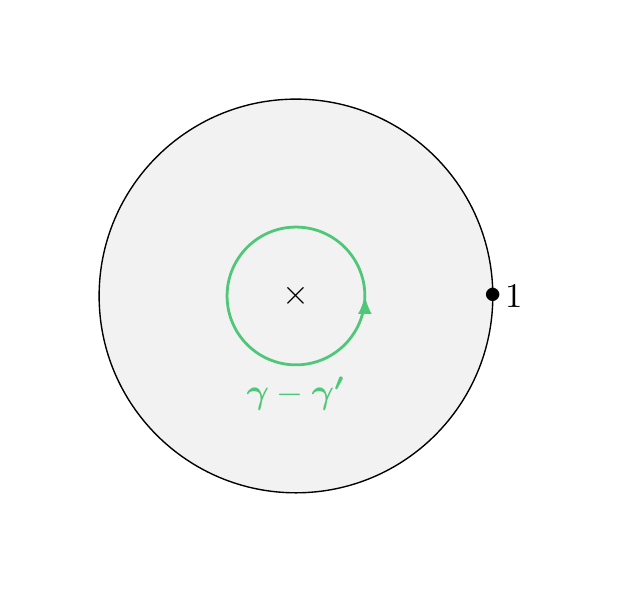}
  \includegraphics[scale=.6]{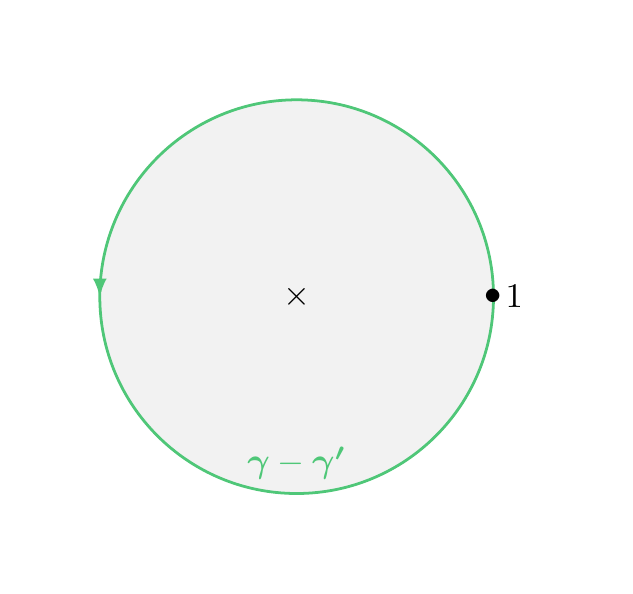}
  \caption{\scriptsize Construction of the map $\Aflat^0 \to {\rm Map}_0(D^*,G)$ via parallel transport along a path. Two paths differ by a parallel transport along a loop winding around the boundary.}
  \label{fig:moduli_space}
\end{figure}

The equations of motion for $X$ are
\begin{equation}
  \begin{alignedat}{2}
    &F_A = 0 \qquad& &\text{(bulk)}, \\
    &A\rvert_{\del D^*} = Xdx\rvert_{\del D^*}  \qquad& &\text{(boundary)}.
  \end{alignedat}
\end{equation}
Hence, integrating out the scalar field $X$, the path integral localizes to the moduli space of flat connection on $D^*$.
To determine the moduli space in question, we first point out that the action is gauge invariant only under those gauge transformations that are trivial on the boundary. 
Indeed, let $\GG = C^{\infty}(D^*,G)$ be the full gauge group, acting on the fields $X$ and $A$ as
\begin{equation}
  X^g = g X g^{-1}, \qquad A^g = g A g^{-1} - dg g^{-1}.
\end{equation}
Then the action is invariant only up to a boundary term:
\begin{equation}
  \delta_g S = S(X^g,A^g) - S(X,A) = - \oint_{\del D^*} \tr X g^{-1}dg.
\end{equation}
However, the normal subgroup $\GG_0 = \{ g \in \GG \mid g\rvert_{\del D^*} = Id \}$ of gauge transformations which are trivial on the boundary leaves the action invariant.
The path integral therefore localizes over the space of flat connections $\Aflat$ modulo the aforementioned gauge transformations: $\mathcal M_0 = \Aflat /\GG_0$.
Flat connections on $D^*$ are characterized by their holonomy around the boundary.
We restrict ourselves to flat connections $A \in \Aflat^0$ whose holonomy around the boundary is the identity.
The moduli space  $\mathcal M_0 = \Aflat^0 /\GG_0$ can be identified with the space of based loops \mbox{$\Omega G = \{ g \in C^{\infty}(S^1,G) \mid g(1) = Id \}$} as follows: let 
\[
  {\rm Maps}_0(D^*,G) = \{ f \colon D^* \to G \mid f(1) = Id \}
\]
be the space of based maps from $D^*$ to $G$. 
We define a map
\begin{equation}
  \begin{split}
    \mathcal M_0 &\to {\rm Maps}_0(D^*,G) / \GG_0 \cong \Omega G \\
    A &\mapsto ( f_A \colon x \mapsto P_A(1 \leadsto x) )
  \end{split}
\end{equation}
where $P_A(1 \leadsto x)$ denotes the parallel transport from $1 \in D^*$ to $x \in D^*$ defined by $A$.
This map is well-defined and in particular independent of the chosen path $1 \leadsto x$. 
Indeed, since the connection is flat, the parallel transport map only depends on the homotopy type of the path.
If now $\gamma, \gamma' \colon 1 \leadsto x$ are any two paths connecting $1$ and $x$, then their difference is a loop (which possibly winds several times around the puncture at the origin), c.f.\ Figure \ref{fig:moduli_space}. 
This loop is always homotopic to a multiple of a loop winding around the boundary.
Since the holonomy of the connection around the boundary is trivial, the parallel transport maps along $\gamma$ and $\gamma'$ are the same. 
The inverse map is constructed as follows: given a based loop $g \in \Omega G$, we can extend it to the punctured disk.
Indeed, if $r\colon D^* \to S^1$ is a retraction, then $\tilde g = g \circ r \colon D^* \to G$ is an extension of the based loop $g$ to $D^*$.
We then get a flat connection by setting $A = -d\tilde g \tilde g^{-1}$. 

The path integral thus localizes over $\Omega G$, with any connection being parametrized by an element $g \in \Omega G$ as $A = -dgg^{-1}$.
Using the boundary equations of motion $A \rvert_{\del D^*} = Xdx \rvert_{\del D^*}$ the boundary action becomes 
\begin{equation}
  S_{\del}(g) = \frac{1}{2} \oint_{S^1} \tr (g'g^{-1})^2 dx
  \label{eq:bdry_action}
\end{equation}
which describes a quantum mechanical free particle moving in the group manifold $\PSL$.
Notice that the boundary action coincides with the Hamiltonian \eqref{eq:H_Omega_G}.

\begin{remark}
It is interesting to compare the above procedure to related constructions in the literature. For example, an action of the type \eqref{eq:bdry_action} is obtained in \cite{Mertens-Turiaci} where a different type of ``puncture'' is considered -- the authors study the BF theory on a full disk in the presence of a defect, coming from a Wilson line in the parent 3-dimensional Chern-Simons theory. This restricts the holonomy of the loop $g$. In our case, we treat the puncture as a boundary, which a priori does not restrict the holonomy. Later we impose such a restriction by hand.
Furthermore, the study of the Schwarzian theory on surfaces of arbitrary topology appeared also in the work \cite{Saad-Shenker-Stanford} of Saad, Schenker and Stanford. However, their focus lies on the (hyperbolic) geometry of the worldsheet, while the present text does not admit a direct gravitational interpretation and thus is not concerned with the worldsheet geometry. 
\end{remark}

\subsection{Edge states and larger gauge groups.}
Recall that due to the presence of the boundary the action was invariant only under the smaller gauge group $\GG_{0} \subset \GG$ of gauge transformations which are trivial on the boundary. 
Following \cite{Gegenberg-Kunstatter-Strobl} and reference therein, one can reinstate the missing gauge degrees of freedom by allowing so-called edge states: let $\Lambda \in C^{\infty}(\del D^*, G)$ which transforms under a gauge transformation as $\Lambda^g = g\Lambda$.
Then, one can add the following extra term to the action:
\begin{equation}
  S(X,A,\Lambda) = S(X,A) + \oint_{\del D^*} \tr Xd\Lambda \Lambda^{-1}. 
  \label{eq:S_edge_states}
\end{equation}
The new action is indeed invariant under arbitrary gauge transformations:
\[
  \delta_g S(X,A,\Lambda) = -\oint_{\del D^*} \tr X g^{-1}dg + \oint_{\del D^*} \tr X g^{-1}dg = 0.
\] 
The action \eqref{eq:bdry_action} admits a global $G$-symmetry which acts on the loops $g$ by conjugation. 
Allowing the edge states $\Lambda$ to take values only in a certain subgroup $H \subset G$ gauges the global symmetry corresponding to the subgroup $H$, i.e. the gauge group is enhanced to $\GG_H =  \{ g \colon D^* \to G \mid g \rvert_{\del D^*} \in H \}$.

Considering the action \eqref{eq:S_edge_states}, integration over the scalar field $X$ imposes the equations of motion
\[
  \begin{alignedat}{2}
    &F_A = 0 \qquad& &\text{(bulk)}, \\
    &A\rvert_{\del D^*} - Xdx\rvert_{\del D^*} + d\Lambda \Lambda^{-1} = 0 \qquad& &\text{(boundary)}.
  \end{alignedat}
\]
Parametrizing $A = -dgg^{-1}$, the action on the boundary becomes 
\begin{align*}
  S_{\del}(g,\Lambda) &= \frac{1}{2} \oint_{\del D^*} \tr (g'g^{-1} - \Lambda'\Lambda^{-1})^2 dx \\
  &= S_{\del}(g) + S_{\del}(\Lambda) - \oint_{\del D^*} \tr \left( \Lambda'\Lambda^{-1} g'g^{-1} \right),
\end{align*}
where $S_{\del}(g)$ is given as in Equation \eqref{eq:bdry_action}.

In particular, if we choose an Iwasawa decomposition $G = NAK$, where $N$ is the group of upper triangular matrices with ones on the diagonal, $A$ diagonal with unit determinant and $K = {\rm SO}(2)$ a compact subgroup, one can restrict the edge state $\Lambda$ to take values only in $N$:
\[
  \Lambda = \mat{1}{\lambda}{0}{1}, \qquad \lambda \in C^{\infty}(S^1,\RR).
\] 
The path integral now localizes over the space of flat connections modulo the augmented gauge group $\GG_N = \{ g \colon D^* \to G \mid g \rvert_{\del D^*} \in N \}$. 
This moduli space can be identified with the space of based loops, modulo the action of based loops in $N$, i.e.\ the space of based loops in $AK \subset G$.
Importantly, one finds $S_{\del}(\Lambda) = 0$.
The boundary action in the presence of edge states then becomes
\[
  S_{\del} = \int_{\del D^*} \left( \frac{1}{2} \tr (g'g^{-1})^2 - \lambda' J_{-}(g) \right)dx, \qquad g \in \Omega(AK) 
\] 
where 
\[
  J_{-}(g) = \tr \mat{0}{1}{0}{0}g'g^{-1}.
\] 
After an integration by parts, $\lambda$ takes over the role of a Lagrangian multiplier which imposes the condition $J_{-}(g) = cst$.
We will see in the next section that the above constraint arises as a first class constraint in the theory without edge states and can be equivalently seen as a Drinfeld-Sokolov reduction of the moduli space $\mathcal M_0 \cong \Omega G$.
Finally, we will show how the constraint leads to Schwarzian quantum mechanics.



\section{Schwarzian action and Drinfeld-Sokolov reduction}\label{sec:QMonPSL}
\subsection{The constrained model}


As before, let $G = \PSL$.
The considerations of the preceding section lead us to the study of quantum mechanics of a free particle moving in $G$ (for compactified time).
It will be convenient for us to consider $LG/G$ rather than $\Omega G$. By abuse of notation, we shall denote an element of the class in $LG/G$ and one of its representatives by the same letter, i.e.\ we will simply write $g$ instead of $[g]$.
The action\footnote{We scale the action in order to make contact with the formulas from Section~\ref{sec:Kac-Moody-Vir-Orbits}.} of the model is given by \eqref{eq:bdry_action}
\begin{equation}
  S(g) = -\frac{k}{4\pi} \tr \oint (g'g^{-1})^2 dx, \qquad g \in LG/G.
  \label{eq:action_OmegaG}
\end{equation}
Under a small variation of the loop, $g \to g + \delta g$ the action changes by
\[
  \delta S \propto  \tr \oint \left( (g'g^{-1})' \delta g g^{-1} \right)dx.
\] 
The equations of motion, therefore, imply that the current $J(g) = g' g^{-1}$ is conserved.
To fix notation, let 
\begin{equation}
  J_{-}(g) =  \tr \mat{0}{1}{0}{0} g'g^{-1}.
  \label{eq:Jminus}
\end{equation}

The action admits a global $G$ symmetry acting by left multiplication on the loops $g \in LG/G$.
We will be interested in gauging part of this symmetry. 
The left multiplication on the loop group $LG$ induces an action of $LG$ on $LG/G$.
We have
\[
  \delta_h S(g) = S(hg) - S(g) = S(h) - \frac{k}{4\pi} \oint \tr\left( h^{-1}h' g'g^{-1} \right) dx.
\] 
In the case that $h$ takes values in the subgroup $N \subset G$ of upper triangular matrices, one finds
\[
  \delta_h S(g) =  \frac{k}{4\pi}  \oint  u \del_x J_{-}(g) dx, \qquad h(x) = \mat{1}{u(x)}{0}{1}.
\] 
Therefore, imposing the first class constraint $J_{-}(g) = 1$, the action acquires a local $LN$ symmetry.

\begin{remark}
  Equivalently, according to the previous section, we could study the model in the presence of edge states taking values in $N \subset G$.
  Upon integration, the edge states impose the above constraint.
\end{remark}

Let us parametrize elements $g \in G$ by an Iwasawa decomposition\footnote{The advantage of the Iwasawa decomposition is that it is global, unlike the Gauss decomposition, which one finds more often in the literature.} of the form 
\begin{equation}
  g = NAK = \mat{1}{F}{0}{1} \mat{a^{-1}}{0}{0}{a} \mat{\cos(\theta/2)}{-\sin(\theta/2)}{\sin(\theta/2)}{\cos(\theta/2)}, 
  \label{eq:Iwasawa}
\end{equation}
where $ F \in \RR,~ a \in (0,\infty),~ \theta \in [0,2\pi]$.
Elements of the loop space $LG$ can now be parametrized in terms of real-valued functions 
\[
  F \colon \RR \to \RR, \qquad a \colon \RR \to (0,\infty), \qquad \theta \colon \RR \to \RR
\]
satisfying the periodicity conditions
\[
  F(x + 2\pi) = F(x), \qquad a(x + 2\pi) = a(x), \qquad \theta(x + 2\pi) = \theta(x) + 2\pi n,
\] 
where $\theta$ is defined up to an overall constant $2\pi k$, i.e.\ we identify \[\theta(x) \sim \theta(x) + 2\pi k.\]
The number $n \in \ZZ$ is called the degree (or winding number) of $\theta$.

In terms of these functions, the action~\eqref{eq:action_OmegaG} is
\begin{equation}
  S(F,a,\theta) 
  = -\frac{k}{2\pi} \oint \left( -\frac{1}{4} \theta'^2 + \left( \frac{a'}{a} \right)^2 + \frac{1}{2} a^2 \theta' F'  \right)dx.
  \label{eq:action_F_a_theta}
\end{equation} 
It is now straightforward to read off the conjugate momentum for $F$:
\begin{equation}
  \pi_F = \frac{1}{2} a^2 \theta'.
\end{equation}
Fixing $\pi_F$ to be a constant imposes a first class constraint:
\begin{equation}
  2\pi_F = 1 \iff a^2\theta' = 1.
  \label{eq:constraint}
\end{equation}
Notice in particular that $\pi_F = J_{-}(g)$.
%
%
%

As we have discussed before, the first class constraint generates a gauge symmetry in $F$.
Indeed, by setting $a^2\theta' = 1$, the last term in \eqref{eq:action_F_a_theta} becomes $\oint F'dx$ and hence vanishes, making the $LN$ symmetry manifest. 
We are therefore interested in the following partition function:
\begin{equation}
  Z^{\rm red} = \int_{LG/G} \frac{d\lambda(g)}{{\rm vol}(LN)}\; \delta( J_{-}(g) - 1/2) e^{-S(g) / \hbar},
  \label{eq:Z_1d_Bershadsky_Ooguri}
\end{equation} 
where the measure $d\lambda(g)$ is taken to be the symplectic measure on $LG/G$.
The above is a one-dimensional analog of the constrained WZW model studied by Bershadsky and Ooguri in \cite{Bershadsky-Ooguri}.
Using the constraint~\eqref{eq:constraint} in the Equation~\eqref{eq:action_F_a_theta}, the action becomes
\begin{equation}
    S(\theta) = -\frac{k}{2\pi} \oint\left(  -\frac{1}{4}\theta'^2 + \frac{1}{4} \left( \frac{\theta''}{\theta'} \right)^2 \right) dx 
    = \frac{k}{4\pi} \oint\left( \frac{1}{2}\theta'^2 +\{ \theta, x \}  \right) dx,
    \label{eq:constrained_action}
\end{equation}
where we used Equation~\eqref{eq:Schwarzian} to write the action explicitly in terms of the Schwarzian derivative.

In the Iwasawa decomposition \eqref{eq:Iwasawa}, the parameter $\theta$ corresponding to an element 
\[g = \mat{\alpha}{\beta}{\gamma}{\delta}\in G,\]
is given by 
\[\tan (\theta/2)=\gamma/\delta.\] 
Thus the right $G$ action on $LG$ descends to the following action on $\theta$'s:
\begin{equation}\label{eq:right_G_action_on_theta}
  \mat{a}{b}{c}{d}\colon \qquad \tan(\theta/2) \quad \mapsto \quad \tan(\widetilde \theta/2)=\frac{a\tan(\theta/2) + c}{b\tan(\theta/2) + d}.
\end{equation}

The constraint~\eqref{eq:constraint} also implies that $\theta' > 0$, i.e.\ $\theta$ is an increasing map and hence the degree $n$ is strictly positive. For any such fixed $n$, let us denote the space of possible $\theta$'s by $L_{(n)}^{+}S^1$. Note that we can assign an element in $L_{(n)}^{+}S^1$ to every oriented diffeomorphism $\phi$ of the circle, simply by taking $n\phi$. It is easy to see that this gives a surjective map
\[\Diff^{+}(S^1)\to L_{(n)}^{+}S^1.\]
On the other hand, every $\theta\in L_{(n)}^{+}S^1$ has $n$ distinct preimages $\phi\in\Diff^{+}(S^1)$, which differ by an integer multiple of $2\pi/n$.
In fact, the above map gives rise to an isomorphism
\[L_{(n)}^{+}S^1\,/\,\PSL\;\cong\; \Diff^{+}(S^1)\,/\,\SLn\;\cong\; \OO_n,\]
where $\OO_n$ is the exceptional Virasoro orbit $\OO_{n}$ passing through $b_0 = -\frac{cn^2}{48\pi}$, c.f.\ \eqref{eq:orbit_n}. 

The partition function thus splits into a sum over topologically distinct sectors which are labeled by the winding number. 
Each sector is governed by an action
\begin{equation}
  S_n(\phi) = \frac{k}{4\pi} \oint \left( \frac{n^2}{2}\phi'^2 + \{ \phi, x \}  \right)dx,
  \label{eq:Sn}
\end{equation}
where $\phi$ is an oriented diffeomorphism of $S^1$, modulo the $\SLn$-action\footnote{In order to match with the usual form of the M\"obius transformations, we now consider the left action corresponding to \eqref{eq:right_G_action_on_theta}, given by $h\colon g\mapsto gh^T$.} 
\begin{equation}
  \tan(n\phi/2) \quad  \mapsto \quad  \tan(n\widetilde\phi/2)=\frac{a\tan(n\phi/2) + b}{c\tan(n\phi/2) + d}.
\end{equation}

Now, the key observation is that the orbits $\OO_n$ are symplectic (see the discussion in Section \ref{sec:Virasoro_orbits}) with symplectic structure
\begin{equation}
  \omega_n =  -\frac{c}{48\pi}\oint n^2 \delta \phi \wedge \delta \phi' - \frac{\delta \phi' \wedge \delta \phi''}{\phi'^2}.
  \label{eq:omega_n}
\end{equation}
Moreover, the actions $S_n$ are the Hamiltonians \eqref{eq:Vir_Hamiltonian} for the $S^1$-action which rotates the loops of $\OO_{n}$. 
A comparison relates the level $k$ with the central charge $c$:
\[
  c = 6k.
\] 
The constrained partition function is therefore the sum over all of the aforementioned topologically distinct sectors:
\begin{equation}
  Z^{\rm red} = \sum_{n = 1}^{\infty} Z_n, \qquad Z_n = \int_{\OO_n} e^{-S_n(\phi)/\hbar}d\lambda_n, 
  \label{eq:Z_red}
\end{equation} 
where $d\lambda_n$ is the symplectic measure of the reduced configuration space $\OO_n$.

Indeed, notice that the symplectic form on $LG/G$ reduces to the symplectic form on $\OO_n$:
Substituting the Iwasawa decomposition~\eqref{eq:Iwasawa} into the symplectic form~\eqref{eq:OmegaG_omega} yields: 
\begin{equation}
  \omega = \frac{k}{4\pi}\oint 2 \frac{\delta a \wedge \delta a'}{a^2} - \frac{\delta \theta \wedge \delta \theta'}{2} +  \delta F \wedge \delta(a^2\theta'), 
\end{equation}
which, after imposing the constraint~\eqref{eq:constraint}, becomes
\begin{equation}
  \omega^{\rm red} = \frac{k}{8\pi}\oint  \frac{\delta \theta' \wedge \delta \theta''}{\theta'^2} - \delta \theta \wedge \delta \theta'.
\end{equation}
With $\theta = n \phi$ and $c = 6k$ we therefore find that the symplectic form restricted to the (totally) reduced configuration space is given by
\begin{equation}
   \omega_n = -\frac{c}{48\pi}\oint n^2 \delta \phi \wedge \delta \phi' - \frac{\delta \phi' \wedge \delta \phi''}{\phi'^2}
\end{equation}
which coincides with the symplectic form on $\OO_n$, c.f.\ \eqref{eq:Vir_omega}.
In this way, the symplectic measure $d\lambda(g)$ on $LG/G$ gives rise to a symplectic measure $d\lambda_n$ on the reduced spaces $\OO_n$.

Finally, returning to \eqref{eq:Zn}, each of the $Z_n$ are integrals over an infinite-dimensional symplectic manifold endowed with a circle action.
In each case, the integrand is the exponential of the Hamiltonian which generates this circle action.
In the case of a finite-dimensional symplectic manifold, this would be precisely the setup suited for Duistermaat-Heckman integration. 
Hence, by analogy, we define the $Z_n$ by the right hand side of the Duistermaat-Heckman integration formula, namely
\begin{equation}
  Z_n = \int_{\OO_n} e^{-S_n/\hbar} d\lambda_n := \sum_{p} \frac{e^{-S_n(p)/\hbar}}{\prod_j \tfrac{1}{2\pi\hbar} w_j(p)},
\end{equation}
with $p$ being the fixed points and $w_j(p)$ the corresponding weights of the $S^{1}$-action.

\subsection{Interpretation in terms of Drinfeld-Sokolov reduction}\label{sec:DS}
At this point, it is instructive to take a step back and to analyze the geometric meaning of the constraint~\eqref{eq:constraint}.
We will see by applying the mechanism of Section \ref{subsec:DS} that the constrained model arises naturally as a Drinfeld-Sokolov reduction of the configuration space $LG/G$. 

We recall that the symplectic form on $LG/G$ is given by
\begin{equation}
  \omega= \frac{k}{4\pi} \tr \oint \left( g^{-1}\delta g \wedge (g^{-1}\delta g)' \right)dx.
  \label{eq:omega_pre_sympl}
\end{equation}
The action~\eqref{eq:action_OmegaG} is then the Hamiltonian for the $S^1$-action which corresponds to rotating the loop:
\[
  S^{1} \times LG/G \to LG/G, \qquad (t,g(x)) \mapsto g(x + t).
\] 
Let now $N \subset G$ be the subgroup 
of upper triangular matrices.
We shall apply the construction of Subsection \ref{subsec:DS} to the corresponding pair of loop groups $LN\subset LG$. Denoting the Lie algebra of $N$ by $\n$, we see that the action of $LN$ on $LG/G$ is Hamiltonian.
The moment map 
\[
  \mu \colon LG/G \to L\n^*
\] 
is given by the projection of $g'g^{-1} \in L\g \cong L\g^*$ onto $L\n^* \cong C^{\infty}(S^1)$. 
%
%
Consider the preimage of any real positive number $q \in \RR_{+} \subset L\n^*$: 
in terms of the Iwasawa decomposition \eqref{eq:Iwasawa}, elements of $\mu^{-1}(q)$ satisfy
\begin{equation}
  g'g^{-1} = \mat{-\tfrac{a'}{a} + \tfrac{1}{2} a^2 \theta' F}{F' + 2\tfrac{a'}{a}F - \tfrac{1}{2}(a^{-2} + a^2F^2)\theta'}{\tfrac{1}{2}a^2\theta'}{\tfrac{a'}{a} - \tfrac{1}{2}a^2\theta'F} = \mat{*}{*}{q}{*}.
  \label{eq:dgg_a_F_theta}
\end{equation} 
which gives the condition
\[
  \frac{1}{2}a^2\theta' = q, \qquad q > 0,
\] 
which, in turn, is equivalent to the constraint~\eqref{eq:constraint}.
Notice that the constraint is fixed under the action of $LN$. 
Thus, the constrained theory, whose space of fields is $\mu^{-1}(q)$, exhibits a gauge symmetry, corresponding to a shift of $F$.
The reduced configuration space is $\mu^{-1}(q)/LN$ which can be seen as the symplectic reduction of $LG/G$ with respect to the aforementioned moment map.
Notably, any class $[g] \in \mu^{-1}(q)/LN$ has a unique representative  
of the form
\[
  \mat{0}{-\tfrac{1}{2q} \left( \tfrac{\theta'^2}{2} + \{ \theta , x \}  \right)}{q}{0}
\] 
which is obtained by using the gauge freedom to set $F = \tfrac{1}{q}\tfrac{a'}{a}$ in \eqref{eq:dgg_a_F_theta} and using the constraint $q = \tfrac{1}{2}a^2\theta'$ to express $a$ in terms of $\theta$.
As before, the constraint imposes $\theta'>0$ and so we can write
\[\mu^{-1}(q)/LN\cong\big\{\tfrac{\theta'^2}{2} + \{ \theta , x \}\mid \theta\in LS^1, \theta'>0\big\}.\]
Using again the parametrization $\theta = n \phi$ for $n>0$ and $\phi \in \Diff^+(S^1)$, we can write
\[\mu^{-1}(q)/LN\cong\coprod_{n\geqslant 1}\big\{\tfrac{n^2}{2}\phi'^2 + \{ \phi , x \}\mid \phi\in \Diff^{+}(S^1)\big\}.\]
Since the transformations of $\phi$'s that preserve the expression $\tfrac{n^2}{2}\phi'^2 + \{ \phi , x \}$ correspond precisely to the $\SLn$ group, we finally obtain
\[\mu^{-1}(q)/LN\cong\coprod_{n\geqslant 1} \Diff^{+}(S^1)/\SLn\cong \coprod_{n\geqslant 1} \OO_n.\]

\subsection{Calculation of the partition function}\label{sec:ZDH}
In order to be self-contained, we now give an explicit calculation of the partition functions $Z_n$.
The final result is already known and was computed by the saddle-point method \cite{Mertens-Turiaci,Stanford-Witten}.
Here, we will apply the formal Duistermaat-Heckman integration formula directly, finding fixed points and weights explicitly.

We recall that $Z_n$ is defined by:
\begin{equation}
  Z_n = \int_{\OO_n} e^{-S_n / \hbar} d\lambda_n := \sum_{p} \frac{e^{-S_n(p) / \hbar}}{\prod_j \tfrac{1}{2\pi\hbar} w_j(p)},
  \label{eq:Zn_standalone}
\end{equation}
where the sum runs over all fixed points $p$ of the $S^1$-action (rotating the loop) on $\OO_n$ and $w_j(p)$ denote the weights (at $p$) of the aforementioned circle action.
We therefore have to compute two things: the fixed points $p$ and the weights $w_j(p)$.

Recall that $\OO_n$ is the left 
quotient of the group $\Diff^{+}(S^1)$ by its subgroup $\SLn$.
The $S^1$-action, which rotates the source circle, corresponds to the right multiplication by $S^1 \subset \SLn$: 
$t\cdot\phi(x) = \phi(x + t)$.
The fixed points on $\OO_n$ correspond therefore to (the cosets of) elements $\phi \in \Diff^{+}(S^1)$, s.t.\ for every $t \in S^1$ there exists $\chi \in \SLn$ satisfying $\phi \circ t =\chi\circ \phi$. 
Equivalently, 
\[
  I_{\phi}(t):=\phi \circ t \circ \phi^{-1} \in \SLn, \qquad  \forall\, t \in S^1.
\]
Since $I_{\phi}$ is a homomorphism, its image $I_{\phi}(S^1)$ is a torus in $\SLn$, and therefore can be brought to the torus $S^1 \subset \SLn$ by conjugation, i.e.\
\[
  g\circ I_{\phi}(S^1)\circ g^{-1} = I_{g\circ \phi}(S^1)\; \subset \;S^1, \qquad \text{for some }g \in \SLn.
\]
Replacing $\phi$ by $g^{-1}\circ \phi$ (they belong to the same equivalence class) we obtain a homomorphism $I_{\phi} \colon S^1 \to S^1$, which implies that
$I_{\phi}(t)=nt$ for some $n \in \ZZ$.

Equivalently,
\[\phi(s+t)=\phi(s)+nt,\quad \forall \,s,t\in S^1.\]
Differentiating with respect to $s$ we see that $\phi'$ is a constant function and thus $\phi$ is necessarily a (constant) rotation. Since rotations belong to the coset of the identity, the only fixed point is the identity $\phi(x) = x$.
\begin{remark}
  The same argument holds for the non-exceptional Virasoro orbits $\OO_b$ passing through $(b,c) \in \Vir^*$ with $b$ constant.
  Hence, the $S^1$-action on $\OO_b$ which rotates the source circle has as well only one fixed point corresponding to the identity.
\end{remark}

To compute the weights of the $S^1$-action on $T_{id}\OO_n$, note that the latter is naturally identified with $\Vect(S^1) /\langle \hat L_0, L^\pm_n \rangle$.
The $S^1$-action is generated by $\hat L_0$. Recall from Equation (\ref{eq:action_L0_on_Lpm}) that $\hat L_0$ acts on the two-dimensional subspaces spanned by $L_m^\pm$ with weight $m$.
Therefore, the denominator in the Duistermaat-Heckman formula \eqref{eq:DH_integration_formula} is given by the infinite product
\[
  \prod_{\substack{ m = 1 \\ m \neq n }}^{\infty} \frac{m}{2\pi\hbar} = \frac{2\pi\hbar}{n} \prod_{m = 1}^{\infty} \frac{m}{2\pi\hbar} = \frac{4\pi^2\hbar^{3/2}}{n},
\] 
which is understood in the zeta-regularized sense.
With 
\[
  S_n(id) = \frac{cn^2}{24}
\] 
the regularized partition function~\eqref{eq:Zn_standalone} in the $n$-th topological sector is therefore given by
\begin{equation}
  Z_n = \frac{n}{4\pi^2\hbar^{3/2}} \exp \left( -\frac{c n^2 }{24 \hbar} \right)
  \label{eq:Zn}
\end{equation}
and hence the constrained partition function is
\begin{equation}
  Z^{\rm red} = \frac{1}{4\pi^2\hbar^{3/2}} \sum_{n \geqslant 1}\, n\, \exp\left( -\frac{cn^2}{24\hbar} \right).
  \label{eq:Z_pathintegral_final}
\end{equation}
Notice that in order for the partition function to be well-defined, we need $c > 0$.
\begin{remark}
  We stress that the above calculation of the partition function should be regarded as a formal application of the Duistermaat-Heckman formula.
  This is neccessary, since apart from the standard issues of working with infinite-dimensional integrals, we encounter a stability problem related to the fact that the fixed point is a saddle point of $S_n$. 
  This was observed in \cite{Alekseev-Shatashvili2,Mertens-Turiaci,Stanford-Witten}.
  Indeed, let us expand the field $\phi$ in a neighborhood of the fixed point,
  \[
    \phi(x) = x + \sum_{m \in \ZZ} \phi_m e^{imx},
  \] 
  where the Fourier modes $\phi_m$ are subject to the reality condition $\bar\phi_m = \phi_{-m}$.
  The action then becomes
  \[
    S_n(\phi) = \frac{cn^2}{24} + \frac{c}{12}\sum_{m = 1}^{\infty} m^2 (n^2 - m^2) \lvert \phi_m \rvert^2 + \OO(\lvert \phi_m \rvert^3),
  \]
  which has an indefinite quadratic part and thus poses a problem for a direct application of the Duistermaat-Heckman integration formula. 
  Nevertheless, the formal calculation presented above seems to yield a meaningful answer. For one thing, it was shown in \cite{Alekseev-Shatashvili2} that \eqref{eq:Zn} determines the correct asymptotic behaviour of Virasoro characters.
  Furthermore, it was conjectured by the same authors that upon a suitable change of integration domain (akin to the steepest descent method), the integral becomes well-defined and reproduces the formal result.
\end{remark}

\begin{remark}\label{rmk:arbitrary_length_of_circle}
  In the literature, one usually starts with the Schwarzian action defined over a circle of length $\beta$, c.f.\ \cite{Mertens-Turiaci,Stanford-Witten}
  \[
    S(\theta)/\hbar = \frac{1}{\hbar}\cdot \frac{k}{4\pi} \int_{0}^{\beta} \left( \frac{1}{2} \theta'^2(y)  + \{ \theta, y \} \right) dy,
  \] 
  where we now allow arbitrary winding numbers, i.e.\ 
  \[
    \theta(x + \beta) = \theta(x) + 2\pi n.
  \] 
  This situation can be related to the calculations presented above by considering the change of coordinates $y = \tfrac{\beta}{2\pi} x$ where $x \in [0,2\pi]$ under which the Schwarzian action becomes 
  \begin{align*}
    S(\theta)/\hbar = \frac{2\pi}{\beta \hbar}\cdot\frac{k}{4\pi} \int_{0}^{2\pi} \left( \frac{1}{2} \tilde\theta'^2(x)  + \{ \tilde\theta, x \} \right) dx, \qquad \tilde\theta(x) = \theta\big( \tfrac{\beta}{2\pi}x\big). 
  \end{align*}
  Therefore, the calculation for arbitrary circumferences $\beta$ can be reduced to the calculations presented above by replacing $\hbar$ by $\hbar \beta / 2\pi$. 
  Therefore, the partition function for the circle of length $\beta$ is given by
  \[
    Z^{\rm red}(\beta) = \frac{1}{(2\pi)^{1/2}(\beta\hbar)^{3/2}} \sum_{n \geqslant 1}\, n\, \exp\left( -\frac{\pi c n^2}{12 \beta \hbar} \right).
  \] 
  This recovers the results from \cite{Mertens-Turiaci,Stanford-Witten} in the subsector $n=1$.
\end{remark}

\subsection{Sources of conserved charges and non-exceptional Virasoro orbits} 
It turns out that one can treat more general actions than the one for a free particle moving in $G$ by the same methods. 
Suppose that we fix a particular edge state\footnote{Strictly speaking, $\Lambda$ is multivalued. However, we are only interested in the corresponding connection $d\Lambda \Lambda^{-1}$, which is well-defined.} $\Lambda$ taking values in $K \subset G$: let $\Lambda = \exp(- x u_0)$ where 
\begin{equation}
  u_0 = \frac{u}{2} \mat{0}{-1}{1}{\phantom{-}0}, 
  \label{eq:v0}
\end{equation}
for a generic real number $u$.
Then, the $BF$ action in the presence of $\Lambda$ becomes  
\[
  S(X,A,u) =  \int_{D^*} \tr X F_A + \oint_{\del D^*} \tr X (A + u_0 dx) - \frac{1}{2} \tr X^2 dx. 
\] 
The boundary equations of motion for $X$ are
\[
  Xdx = A + u_0dx.
\] 
Integrating out $X$, the path integral localizes over the moduli space of flat connections and imposing the unit holonomy condition, we can parametrize $A$ by $A = g^{-1}dg$. The boundary action thus becomes
\[
  S(A,u) = \frac{1}{2} \oint \tr (g^{-1}g' + u_0)^2 dx = - \frac{\pi u^2}{2} + \oint  \left( \frac{1}{2} \tr (g^{-1}g')^2 - \frac{u}{2}  J_K(g^{-1})  \right) \!dx  
\] 
where
\[
  J_K(g) = \tr \mat{0}{-1}{1}{0} g'g^{-1}
\] 
is the current corresponding to the compact subgroup $K$. 
In particular, $u$ plays the role of a source for the charge 
\[
  Q_K(g) = \oint J_K(g^{-1})dx.
\] 
Indeed, taking derivatives of the partition function with respect to $u$, one obtains correlation functions of powers of the charge.
For example, one has
\[
  2\hbar\frac{\del Z(u)}{\del u}\Big\rvert_{u=0} = \int d\mu(g)\; e^{-S(g)/\hbar} Q_K(g) = \left\langle Q_K(g) \right\rangle.
\] 

\begin{remark}
  By choosing more general $u_0$, one obtains an action in the presence of sources for more general conserved charges. 
\end{remark}

Motivated by the above considerations and in order to make contact with the Example \ref{ex:LG/K}, let us now fix 
\begin{equation*}
  u_0=\frac{2\pi}{k}v_0 = \frac{\pi v}{k} \mat{0}{-1}{1}{\phantom{-}0}, \qquad v>0.
\end{equation*}
Consider the action
\begin{equation}
  S(g) = -\frac{k}{4\pi} \oint \left( g^{-1}g' + \frac{2\pi}{k}v_0 \right)^2 dx = -\frac{k}{4\pi} \tr \oint (h'h^{-1})^2dx,
  \label{eq:Sv}
\end{equation}
where we introduced the quasi-periodic element 
\[
  h = g\exp( \tfrac{2\pi}{k} v_0 x).
\] 

For any element $g \in LG$, we have an Iwasawa decomposition for $h$ (c.f~\eqref{eq:Iwasawa}):
\begin{equation}
  h = \mat{1}{F}{0}{1} \mat{a^{-1}}{0}{0}{a} \mat{\cos(\tilde\theta/2)}{-\sin(\tilde\theta/2)}{\sin(\tilde\theta/2)}{\phantom{-}\cos(\tilde\theta/2)}, \qquad \tilde\theta = \theta + \frac{2\pi v x}{k},
\end{equation}
where $F$, $a$ and $\theta$ are, as before, real-valued periodic (respectively quasi-periodic) functions on $\RR$.
Note that if $\theta$ has winding number $n \in \ZZ$, then $\tilde\theta$ satisfies the quasi-periodicity condition
\begin{equation}
  \tilde\theta(x + 2\pi) = \tilde\theta(x) + 2\pi \left( n + \frac{2\pi v}{k} \right).
  \label{eq:periodicity_tilde_theta}
\end{equation}

Proceeding as in the previous section, we impose the first class constraint $a^2\tilde\theta' = 1$.
In particular, $\tilde\theta$ is an increasing map, $\tilde\theta' > 0$.
Equation \eqref{eq:periodicity_tilde_theta} therefore implies that $n + 2\pi v / k > 0$.
Any such map can be parametrized by an orientation preserving diffeomorphism $\phi \in \Diff^{+}(S^1)$:
\begin{equation}
  \tilde\theta(x) = 
    \left( n + \frac{2\pi v}{k} \right) \phi(x) \equiv \kappa_n(v)\phi(x).
  \label{eq:para_tilde_theta}
\end{equation}
Since $\phi(x+2\pi) = \phi(x) + 2\pi$, $\tilde\theta$ satisfies indeed the correct periodicity condition~\eqref{eq:periodicity_tilde_theta}.
In fact, since $\tilde\theta' > 0$, only those $\tilde\theta$ with $\kappa_n(v) > 0$ will contribute to the constrained path integral.
\begin{remark}
  Let us point out that the path integral reduces to an integral over the Kac-Moody orbit $LG/K$, c.f.\ Example \ref{ex:LG/K}.
  As before, the action coincides with the Hamiltonian generating the $S^1$-action on the orbit.
  Again, the constraint $a^2\tilde\theta = 1$ corresponds to a Drinfeld-Sokolov reduction for the $LN$-action on $LG/K$.
\end{remark}

In terms of the fields $(F,a,\tilde \theta)$, the action reads:
\begin{equation}
  S(g,v_0) = -\frac{k}{2\pi} \oint \left( -\frac{1}{4} \tilde\theta'^2 + \left( \frac{a'}{a} \right)^2 + \frac{1}{2} a^2 \tilde\theta' F'  \right)dx.
\end{equation}
%
After imposing the constraint, the partition function splits again into distinct sectors, each governed by an action of the form
\begin{equation}\label{eq:non_exceptional_schwarzian_action}
  S_{n}(\phi,v) = \frac{k}{4\pi} \oint \left( \frac{\kappa_n^2}{2} \phi'^2 + \{ \phi,x \}  \right)dx.
\end{equation}
Now, there exists a residual $S^1$ symmetry which acts by constant shifts of $\phi$: $\phi(x) \to \phi(x) + t$, for $t \in S^1$.
Hence, the (totally) reduced configuration space is isomorphic to the more general Virasoro orbit passing through the point $b_0 =  - c \kappa_n^2 / 48 \pi$, which is isomorphic to $\Diff^{+}(S^1) / S^1$. 
The reduced configuration space is therefore again a symplectic space.
Setting $c = 6k$, the actions $S_{n}$ again coincide with the Hamiltonians generating the $S^1$-action (which rotates the source circle) on the aforementioned Virasoro coadjoint orbits.

By the same arguments as in the previous section, the partition function splits into a sum of integrals over these more general Virasoro orbits which again are defined by the right hand side of the Duistermaat-Heckman integration formula~\eqref{eq:DH_integration_formula}:
\[
  Z_{v}^{\rm red} = \sum_{n + 2\pi v/k>0} \int_{\Diff^{+}(S^1) / S^1} e^{-S_{n}(\phi,v)/\hbar} = \sum_{n + 2\pi v/k>0} \sum_{p} \frac{e^{-S_{n}(p,v)/\hbar}}{\prod_j \tfrac{1}{2\pi\hbar} w_j(p)} 
\] 
where the sum runs over all fixed points $p$ and the $w_j(p)$ are the weights of the $S^1$-action on the tangent space at $p$. 
As discussed previously, there is only one fixed point, namely $\phi = id$ at which the action takes the value $S(id) = \tfrac{c}{24}\kappa_n^2$.
By the same argument as before,  
%
the denominator is given by the zeta-regularized product
\[
  \prod_{m = 1}^{\infty} \frac{m}{2\pi\hbar}
  = 2\pi\sqrt{\hbar}.
\] 
Therefore, the partition function is 
\begin{equation}
Z_{v}^{\rm red} =  \frac{1}{2\pi\sqrt{\hbar}} \sum_{n + 12\pi v/ c > 0}  \exp\left( -\frac{c(n + 12\pi v /c)^2}{24\hbar}  \right).
\end{equation}

\begin{remark}
  It is intriguing that the limit of taking $v$ to zero does not give back the constrained partition function $Z^{\rm red}$. 
  On the other hand, in this limit we recognize that $(\Diff^{+}(S^1)/S^{1}, \omega)$ is only a pre-symplectic space.
  The kernel of $\omega$, in the $n$-th topological sector, is generated precisely by the Virasoro vector fields $L^\pm_n$.
  It is the emergence of these zero modes, which spoils the naive limit of the partition functions. 
  In fact, we can fix the zero modes by considering an additional reduction, namely by taking the quotient with respect to $\ker \omega$. 
  The resulting spaces are exactly the exceptional Virasoro orbits $\OO_n \cong \Diff^{+}(S^1) / \SLn$, whose partition function has been calculated in Equation~\eqref{eq:Zn}.
\end{remark}

\begin{remark}
  As in the case of the exceptional Virasoro orbits, the calculation presented above was carried out for a circle of length $2\pi$. 
  Analogously to the discussion in Remark \ref{rmk:arbitrary_length_of_circle}, the partition function defined by the action \eqref{eq:non_exceptional_schwarzian_action} for a circle of arbitrary length $\beta$ can be obtained from the substitution $\hbar \to \beta\hbar/2\pi$.
  We then find
  \[
    Z_{v}^{\rm red}(\beta) = \frac{1}{\sqrt{2\pi\beta\hbar}} \sum_{n + 12\pi v/ c > 0}  \exp\left( -\frac{\pi c(n + 12\pi v /c)^2}{12\beta\hbar}  \right),
  \]
  recovering the result of \cite{Mertens-Turiaci}.
\end{remark}

\end{document}